# Optimizing Differential Identifiability Improves Predictive Modeling of Cognitive Deficits from Functional Connectivity in Alzheimer's Disease

Diana O. Svaldi PhD[1*], Joaquín Goñi PhD[2,3,4], Kausar Abbas PhD[2,3], Enrico Amico PhD[2,3], David G. Clark[1], Charanya Muralidharan[1], Mario Dzemidzic PhD[1], John D. West MS[1], Shannon L. Risacher PhD[1], Andrew J. Saykin PsyD[1], Liana G. Apostolova MD, MSc[1], for the Alzheimer's Disease Neuroimaging Initiative**

[1] Indiana University School of Medicine, Indianapolis, IN, USA
[2] School of Industrial Engineering, Purdue University, West-Lafayette, IN, USA
[3] Purdue Institute for Integrative Neuroscience, Purdue University, West-Lafayette, IN, USA
[4] Weldon School of Biomedical Engineering, Purdue University, West-Lafayette, IN, USA

[*] **corresponding author:** Diana O. Svaldi PhD, dosvaldi@iu.edu
**Data used in preparation of this article were obtained from the Alzheimer's Disease Neuroimaging Initiative (ADNI) database (adni.loni.usc.edu). As such, the investigators within the ADNI contributed to the design and implementation of ADNI and/or provided data but did not participate in analysis or writing of this report. A complete listing of ADNI investigators can be found at: http://adni.loni.usc.edu/wp-content/uploads/how_to_apply/ADNI_Acknowledgement_List.pdf

**Abstract**

Functional connectivity, as estimated using resting state functional MRI, has shown potential in bridging the gap between pathophysiology and cognition. However, clinical use of functional connectivity biomarkers is impeded by unreliable estimates of individual functional connectomes and lack of generalizability of models predicting cognitive outcomes from connectivity. To address these issues, we combine the frameworks of ***connectome predictive modeling*** and ***differential identifiability***. Using the combined framework, we show that enhancing the individual fingerprint of resting state functional connectomes leads to robust identification of functional networks associated to cognitive outcomes and also improves prediction of cognitive outcomes from functional connectomes. Using a comprehensive spectrum of cognitive outcomes associated to Alzheimer's disease (AD), we identify and characterize functional networks associated to specific cognitive deficits exhibited in AD. This combined framework is an important step in making individual level predictions of cognition from resting state functional connectomes and in understanding the relationship between cognition and connectivity.

**Keywords:** Alzheimer's Disease, AD, cognition, resting state, fMRI, functional connectivity, functional fingerprinting, predictive modeling

## 1 Introduction

The biological underpinnings of all neurodegenerative disorders remain poorly understood, contributing significantly to the bottleneck in treating these disorders [1]. In recent years, the application of analyses based on complex systems approaches for understanding how neural activity and connectivity facilitate cognition has led to significant strides in characterizing these disorders [2, 3]. One such approach, functional brain connectomics, models functional brain networks as pairwise statistical dependencies in regional neural activity. This provides a framework to assess critical aspects of the brain, such as integration and segregation [4], and ultimately communication [5, 6]. At the same time, the advent of functional MRI (fMRI) has allowed for in-vivo characterization of whole brain functional connectomes (FC) in humans [5], leading to the discovery of several critical brain networks implicated in schizophrenia, attention deficit hyperactivity disorder, autism, and Alzheimer's Disease (AD) [3].

Despite their potential to enhance our understanding of neurologic disorders, approaches based on functional connectivity have not yet been used translationally in the treatment of cognitive and behavioral disorders [7, 8]. To advance the treatment of such disorders, there is a critical need to develop clinical biomarkers that are (1) robustly modulated by disease mechanisms and (2) specifically associated with disease related outcomes [8]. Though functional connectivity has shown potential in bridging the gap between pathophysiology and cognition, its clinical use is impeded by unreliable estimation of subject level FC [9], lack of precision in inter-subject differentiability in FC [10], and lack of generalizability of models predicting subject-level cognitive outcomes from FC [8]. Here we show that improving the subject level fingerprint of resting-state FC also improves prediction of a heterogeneous set of cognitive deficits in AD, both in new data from the training cohort as well as data from a validation cohort. We also identify functional networks associated to specific cognitive deficits exhibited in AD.

### 1.1 Towards Improving Clinical Utility of FC

While FC shows differential group level associations across cognitive outcomes [11, 12] and across disease conditions [2, 3, 13-16], it falls short of predicting clinically meaningful outcomes at the individual level. The reason for this, is insufficient “fingerprint” or within-subject reliability and between-subject differentiability to capture individual differences that may be related to cognition or behavior [12, 17-21]. In terms of reliability of FC, it has been shown that reasonable reliability can be achieved with sufficient scan length and that this reliability can be improved when multiple sessions of FC are used [10, 22, 23]. Studies have also shown that FC reliability is different across the brain, with larger cortical nodes displaying the most reliability and within network connections exhibiting greater reliability than between network connections [10]. Several studies have also shown that frequently used 6 min fMRI acquisitions do not have adequate reliability at the edge level, posing significant issues on numerous available clinical fMRI datasets when performing subject level associations. In terms of inter-subject differentiability, recent efforts have shown that individuals can be reasonably distinguished from each other using FC, as measured by identification rate [12, 17, 24], perfect separability rate [10, 17, 24], or differential identifiability [12]. Furthermore, it has been shown that individual level fingerprinting improves with longer scan length [10, 12] and when subjects are performing specific tasks [24]. Finally, there is evidence that FC fingerprint is reduced in individuals with neurologic or psychiatric conditions [13, 25, 26], making association of FC with disease related phenotypes more difficult.

Evidence for fingerprint in FC has opened the door for efforts to improve prediction of cognition and behavior from FC [7, 8, 27-29]. Predictive pipelines typically involve: (1) feature reduction to find FC features that are associated with specific cognitive outcomes, (2) training of a predictive model using these features to predict cognitive outcomes, and (3) evaluation of the accuracy and generalizability of resulting models on external data. It has been demonstrated that using multiple connectomes across sessions or connectomes from different tasks improves predictive power [30]. However, how test/retest reliability of FC features affects their contribution to predicting cognition or behavior is still under debate. One study found no association between the test/retest reliability of a given functional edge and the predictive value of the edge in external data [10]. However, another study [12] showed increased prediction accuracy when edges were chosen based on correlation to cognitive outcome and test/retest reliability, after using PCA to optimize differential identifiability to uncover FC fingerprints.

Though strategies such as increased scan length, multiple acquisitions and adding task-based fMRI may be useful for improving prediction of relevant outcomes, they may not be feasible in clinical settings. Reasons include the associated cost of increased scan time, diminished tolerance in patient populations to long scan times, and impaired ability of patients in performing tasks. Additionally, there are numerous available datasets [16, 31-34] with acquisition protocols that do not have adequate reliability [10] for subject level prediction. Finally, prediction of cognitive outcomes in neurologic and psychiatric populations remains a challenge as these subjects appear to have a reduced fingerprint [25, 26]. To address such issues, Amico and Goni proposed the *differential identifiability* framework ($\mathbb{I}f$) [12], which is a PCA based denoising algorithm to uncover fingerprints in FC and improve between-subject differentiability at the same time. Using data from 100 unrelated subjects in the Human Connectome Project, they demonstrated improvements in FC fingerprint beyond what could be achieved by increasing scan length [12]. This improvement was also observed in FC data from the Alzheimer’s Disease Neuroimaging Initiative (ADNI), a dataset with more traditional acquisition consisting of 140 volumes (7 min scan) split in half to mimic a test/retest setting [13]. Thus, the $\mathbb{I}f$ framework demonstrated improvements in fingerprinting in a “traditional” acquisition performed on a clinical population. However, there is still conflicting evidence on whether increasing FC fingerprint subsequently improves prediction of clinically relevant outcomes. It is important to note that the abovementioned studies specifically assessed whether the test/retest reliability of a functional edge affected the predictive value of that edge in external subjects. Hence, two important questions remain open (1) the level of agreement in feature selection between test and retest data from the same subjects and (2) whether a model built on test data would generalize to re-test data from the same subjects. These two questions are critical, since good performance of predictive models in a test/retest setting is a minimum standard that should be met before testing on external data. Lack of agreement in feature selection between test and retest data indicates a model that overfits the training data and is not generalizable, even to a new session of the same training subjects. Even if this model is somewhat generalizable to external subjects, if it lacks test-retest agreement in feature selection, the model is likely overparametrized and selecting arbitrary features.

In this work, we test the effect of the differential identifiability framework on key properties of models predicting cognitive outcomes related to AD from FC data. We assess performance of the models in both a test/retest setting and in generalization to validation data. When choosing key properties to assess the quality of predictive models for the purposes of predicting and understanding cognitive associations to the brain, it is important to keep in mind that interpretation of anatomical locations of the cognitive correlates of FC are as relevant as the accuracy of prediction. Hence, confirming robustness in the identification of FC features should precede model fitting and assessments of model accuracy. Further, it is important to note that the robustness of both feature selection and coefficient estimation can significantly influence model accuracy and generalizability. Therefore, we propose to evaluate three critical properties for *well-behaved* FC-based predictive models: (1) stability of feature selection in a test/retest setting, (2) specificity of edge selection, and (3) generalizability of the prediction to new data from the same subjects and to validation data.

### 1.2 Opportunities in Alzheimer's Disease

We chose to evaluate these effects in data from the ADNI2, which consists of subjects spanning the AD spectrum . The heterogeneous, gradual progression of cognitive deficits in AD is particularly amenable to study the quality of models predicting cognition from FC. Briefly, in the prodromal AD stage of mild cognitive impairment (MCI) subjects typically manifest episodic memory decline, which is later accompanied by subtle deficits in other domains, and ultimately results in progressive functional impairment as the subject transitions through the mild, moderate and severe stages of dementia [35-38]. Within the AD spectrum there is much individual heterogeneity in terms of disease presentation and progression over time [38], making predictive modeling at the subject level important.

The association between FC changes and cognitive deficits in AD has been subject of intense study to date [16, 39, 40]. Changes in functional networks, primarily the default mode and frontoparietal networks, have been consistently replicated between diagnostic groups [41-43]. Recent studies indicate that FC data can predict subject level diagnostic status [44] and global cognitive decline [45] with reasonable accuracy. Several studies also show relationships between FC data and deficits in specific cognitive domains associated with AD [16, 39, 46].

In this work, beyond assessing group level associations to specific cognitive domains or individual level prediction of cognitive status (impaired versus non-impaired), we present a framework that improves the ability of FC to predict subject level deficits from different cognitive domains. This additionally enables us to assess which RSNs are globally associated to cognition in AD versus RSNs associated to specific deficits observed in AD.

## 2 Methods

### 2.1 Subject Demographics and Cognitive Performance

Data used in the preparation of this article were obtained from the ADNI database (adni.loni.usc.edu). The ADNI was launched in 2003 as a public-private partnership, led by Principal Investigator Michael W. Weiner,

MD. The primary goal of ADNI has been to test whether serial magnetic resonance imaging (MRI), positron emission tomography (PET), other biological markers, and clinical and neuropsychological assessment can be combined to measure the progression of MCI and AD. For up-to-date information, see www.adni-info.org.

In this work, resting state fMRI and neurocognitive testing data from the second phase of the Alzheimer's Disease Neuroimaging Initiative (ADNI2/GO) were used. Our analyses included 82 participants from the original 164 ADNI2/GO individuals with resting state fMRI scans. Subjects were excluded if (1) their amyloid status was not available, (2) were cognitively impaired, but showed no evidence of amyloid-beta (Aß) deposition, and/or had (3) over 30% of fMRI time points censored due to artifacts or head motion (see Section 2.2 for details). Aß status was determined using either mean PET Florbetapir standard uptake value ratio cutoff (Florbetapir > 1.1; UC Berkeley) or CSF Aß level ≤192 pg/mls [5]. The rationale for excluding Aß- cognitively impaired participants was to avoid confounding by non-AD neurodegenerative pathologies. Subjects were stratified into five categories based on their diagnosis and Aß status: (1) Aß- cognitively normal individuals ($\text{CN}_{\text{Aß-}}$, n = 15), (2) Aß+ CN or pre-clinical AD ($\text{CN}_{\text{Aß+}}$, n = 12), (3) early MCI due to AD ($\text{EMCI}_{\text{Aß+}}$, n = 22), (4) late MCI due to AD ($\text{LMCI}_{\text{Aß+}}$, n = 12), and (5) AD dementia ($\text{AD}_{\text{Aß+}}$, n = 21).

We selected five outcome measures for predictive modeling from the ADNI2/GO neurocognitive battery, (www.adni-info.org for protocols): the auditory verbal learning test (AVLT) immediate recall, AVLT delayed recall, clock drawing, Trails B, Animal Fluency. Additionally, the Montreal cognitive assessment (MOCA) was also included as a representative clinical measure of global cognition. All six outcomes selected exhibited a significant diagnostic group effect (see Table1, ANOVA $p<0.05$). Of note, all outcome measures were z-scored, relative to the training data, prior to predictive modeling to allow for direct comparison between models across outcome measures.

**Table 1.** Demographics and Neurocognitive Comparisons of Diagnostic Groups.

| Variable Mean (SD) | $CN_{Aß-}$ (n = 15) | $CN_{Aß+}$ (n = 12) | $EMCI_{Aß+}$ (n = 22) | $LMCI_{Aß+}$ (n = 12) | $AD_{Aß+}$ (n = 21) |
|---|---|---|---|---|---|
| **Age (Years)** | 74.2 (8.8) | 75.9 (7.0) | 72.6 (5.2) | 73.3 (6.1) | 73.5 (7.6) |
| **Sex (% F)** | 64.2 | 41.7 | 50 | 61.6 | 42.9 |
| **Years of Education** | 16.7 (2.3) | 15.8 (2.6) | 15.2 (2.6) | 16 (1.8) | 15.4 (2.6) |
| **MOCA *** | 26.2 (2.6) | 25.3 (2.9) | 22.3 (4.5) | 20.6 (7.1) | 13.4 (5.2) |
| **Auditory Verbal Learning Immediate Recall *** | 11.1 (3.0) | 11.33 (2.9) | 9.9 (3.0) | 7.6 (2.4) | 4.3 (1.6) |
| **Auditory Verbal Learning Delayed Recall *** | 6.2 (4.3) | 7.8 (3.8) | 4.3 (4.0) | 2.8 (2.8) | 0.4 (0.9) |
| **Boston Naming*** | 28.2 (2.0) | 28.7 (1.1) | 27.1 (3.1) | 25.9 (5.0) | 22.4 (6.4) |
| **Animal Fluency *** | 21.1 (3.64) | 20.1 (3.6) | 18.8 (4.2) | 17.4 (4.8) | 12.3 (5.0) |
| **Clock Drawing *** | 4.8 (0.4) | 4.5 (1.0) | 4.6 (0.5) | 3.8 (1.3) | 3.1 (1.3) |
| **Trail Making B *** | 69.0 (22.6) | 81.4 (19.6) | 99.9 (43.1) | 131 (89.0) | 216.9 (75.6) |

* Significant group effect (Chi-squared or ANOVA as appropriate, $\alpha$ = 0.05). Values in parenthesis denote standard deviation

### 2.2 fMRI Data Processing

We used T1-weighted MPRAGE scans and EPI fMRI scans from the initial visit in ADNI2/GO (Philips Platforms, TR/TE = 3000/30ms, 140 volumes, 3.3 mm thickness, see www.adni-info.org for detailed protocols) for estimation of whole-brain FCs. fMRI scans were processed with an in-house MATLAB and FSL based pipeline as described in detail in Amico et al. [32]. This pipeline mainly follows guidelines by Power et al. [47, 48]. For purposes of evaluating reproducibility, we split the processed fMRI time series into halves (mimicking test and retest) and assigned each half for each subject as “restA” or “restB” randomly to avoid biases related to first versus second half of the scan. It is noteworthy that splitting an fMRI session mimics the most ideal test-retest scenario where all conditions are maintained as homogenously as possible.

We obtained two FC matrices from the restA and restB halves of the fMRI time-series for each subject. FC nodes were defined using a 278 region cortical parcellation [49], as detailed in Amico et al. [32], with a modified subcortical parcellation [50], for a total of 286 gray matter regions. We estimated single session FC matrices by calculating the Pearson correlation coefficient [51] between the fMRI time-series of each pair of brain regions. Each region’s time-series was obtained by averaging time-series of all voxels assigned to that brain region. Regions were assigned to one of the seven cortical resting state subnetworks (RSN/RSNs): visual (VIS), somato-motor (SM), dorsal attention (DA), salience (SAL), limbic (L), executive control (EC), and default mode network (DMN) [52], with the remaining regions assigned to a subcortical (SUB) or cerebellar (CER) networks.

## 2.3 Differential Identifiability Framework ($\mathbb{I}f$)

We applied $\mathbb{I}f$ which uses group level principal component analysis (PCA) [53] to find the optimal FC reconstruction point for simultaneous optimization of restA and restB FC reproducibility and between-subject differentiability, measured using differential identifiability ($I_{diff}$, Fig.1) [12]. The "identifiability matrix" ***I*** was defined as the matrix of pairwise correlations (square, non-symmetric) between the subjects' $FC_{restA}$ and $FC_{restB}$. The dimension of ***I*** is $N^2$ where $N$ is the number of subjects in the cohort. Self-identifiability, ($I_{self}$, *Eq. 1*), was defined as the average of the main diagonal elements of ***I***, consisting of correlations between $FC_{restA}$ and $FC_{restB}$ from the same subjects. $I_{others}$ (*Eq. 2*), was defined as average of the off-diagonal elements of matrix ***I***, consisting of correlations between $FC_{restA}$ and $FC_{restB}$ of different subjects. Differential identifiability ($I_{diff}$, *Eq. 3*) was defined as the difference between $I_{self}$ and $I_{others}$.

$$I_{self} = \frac{1}{N}\sum_{i=1}^{N} I_{i,i} \quad (Eq. 1)$$

$$I_{others} = \frac{1}{2\binom{N}{2}}\sum_{i \neq j} I_{i,j} \quad (Eq. 2)$$

$$I_{diff} = 100 * \left(I_{self} - I_{others}\right) \quad (Eq. 3)$$

We applied group level PCA in the FC domain, on a data matrix (Fig.1A-B) containing vectorized $FC_{restA}$ and $FC_{restB}$ (upper triangular of FC matrices excluding main diagonal) from all subjects. Following PCA decomposition (Fig.1C), we iteratively reconstructed (Fig.1D) all FCs and quantified $I_{diff}$ for a range of number of PCs (Fig.1E). FC matrices were reconstructed using the number of PCs optimizing $I_{diff}$. The fingerprint at the functional edgewise level for each subject was evaluated for the original FC matrices using the intraclass correlation coefficient (ICC 2,1) [54].

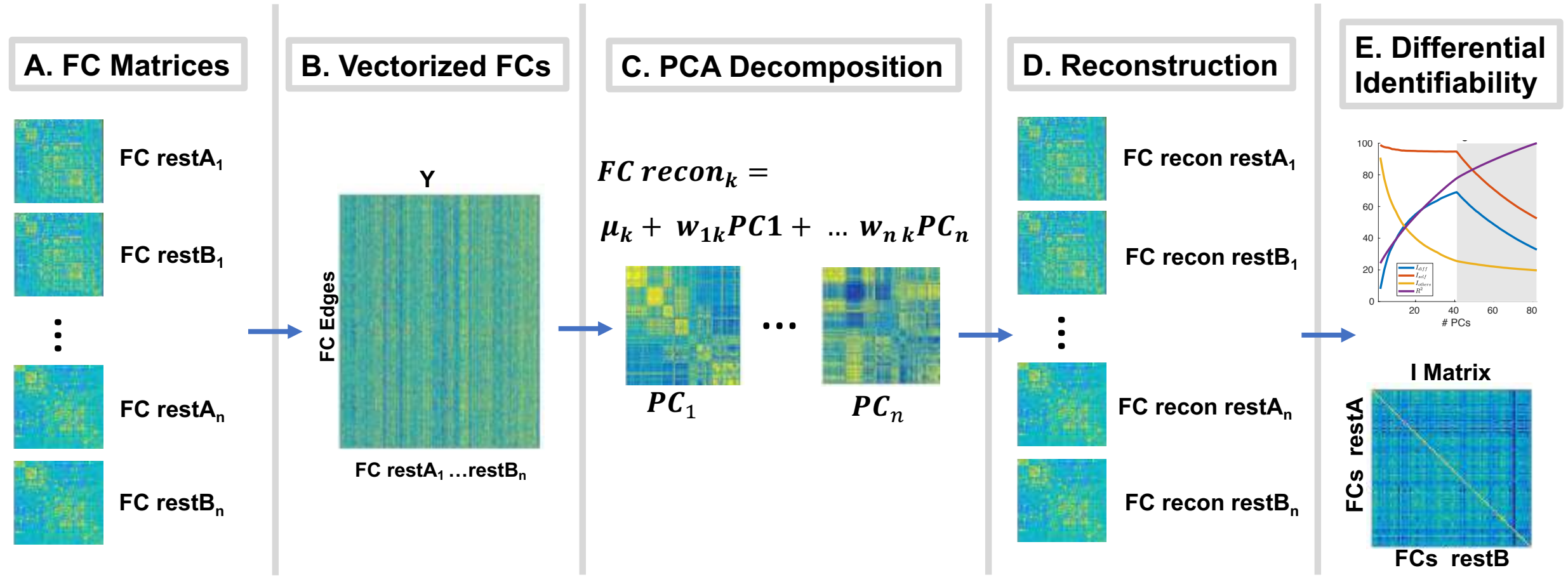

**Figure 1. Differential identifiability framework ($\mathbb{I}f$). (A)** For each subject, two FC matrices (restA and restB) were estimated for each half of the fMRI time-series. **(B)** FC matrices were vectorized (upper triangular) and placed into a group FC matrix. **(C)** PCA decomposition was performed on the group FC matrix. Each PC can be arranged as a matrix in the FC domain. **(D)** Individual FCs were reconstructed using different number of PCs. **(E)** $I_{diff}$ was estimated for different number of PCs (in order of explained variance) and the number of PCs maximizing $I_{diff}$ found.

## 2.4 Assessment of Differential Identifiability Pipeline on Connectome Predictive Modeling

The connectome predictive modeling pipeline (CPM) [27] was used to assess the effect of $\mathbb{I}f$ [12] on predictive modeling of the aforementioned outcome measures. The entire cohort (N=82) was split into training cohort (N=41) and validation cohort (N=41), in a split half, k-fold cross validation fashion as suggested by [22]. In each fold, 41 subjects were randomly selected as training subjects and the other 41 were selected as validation subjects. $\mathbb{I}f$ was performed separately on training versus validation subjects. The training cohort was used in a two-step framework (Fig. 2) to estimate models predicting each outcome measure from FC (Fig. 2A).

The stability and specificity of the predictive modeling pipeline was first evaluated in a test-retest setting on the training subjects. Stability in FC-outcome correlation: For each fold, we computed the correlation between each edge (from a total of 40,755 edges) and each outcome measure. Edges exhibiting an absolute value of correlation above 0.1 were selected to create a positive correlation mask and negative correlation mask. We evaluated stability in edgewise correlation using the Frobenius norm between restA and restB correlations, where values close to zero denote high similarity between restA and restB correlation vectors. Stability in edges selection: We evaluated the similarity of overlap in selected edges as the number of edges selected (same sign) using both restA and restB FCs divided by the number of edges selected using either restA or restB FCs. Specificity of edge selection: Additionally, we evaluated specificity of edge selection by calculating average, pairwise Frobenius norms and mask overlaps across all outcome measures using only restA FCs. Averaged Frobenius norm and percent overlap over the K folds was reported for both stability and specificity across the range of PCs.

Following edge selection, model fitting portion of the CPM pipeline was then performed on restA data and subsequently evaluated on restB data. To estimate a model for each outcome measure, strength (sum of all edges in the mask) in the positive and negative masks were used as predictors in a linear regression model. Mean squared error (MSE) and Pearson correlation of estimated versus observed outcomes were used to evaluate model generalizability from restA to restB data in training subjects. Average MSE and Pearson correlation over the K folds are reported for restA FCs and restB FCs reconstructed across the range of PCs.

Finally, we validated performance of the model on the testing subjects in original FCs versus FCs optimally reconstructed for $I_{diff}$. For estimation of a final model, restA and restB training FCs were averaged. Using strength in the masks from restA training data as predictors, a final model was fit on the restA-restB averaged FCs from the training data. The final model for each outcome was then used to estimate the corresponding outcome measure on restA-restB averaged FC data from the validation subjects. MSE and Pearson correlation were again used to evaluate performance of the model on the test subjects. Averaged MSE and Pearson correlation over the K folds are reported. To compare performance of models built/tested on original FCs versus models build/tested on optimally reconstructed FCs paired t-tests ($\alpha$ = 0.05), using original FCs versus optimal FCs at each fold as paired samples, were conducted on the Pearson correlations.

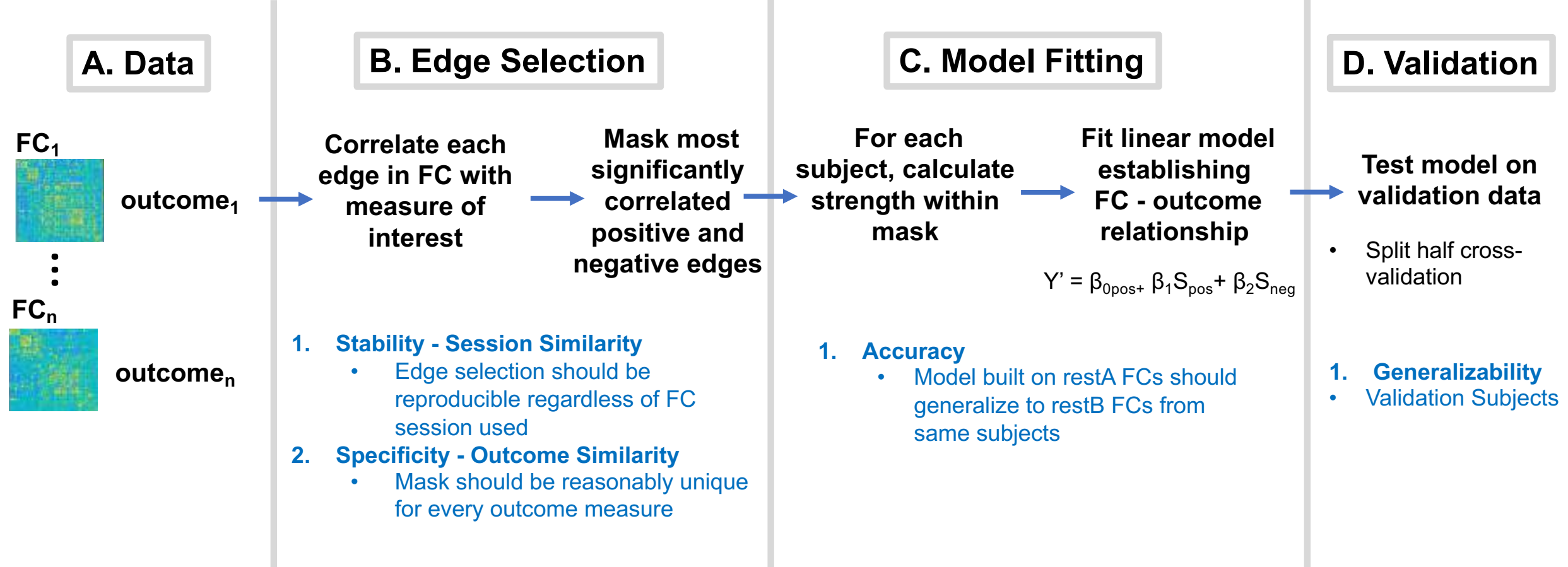

**Figure 2.** Connectome predictive modeling scheme (adapted from [27]). Black text delineates procedures for each step while blue text delineates properties that are important at each step to achieve an overall robust model. **(A)** The goal is to predict the outcome measure from FC data. **(B)** Edgewise correlations were performed with outcome of interest. Most significantly positively and negatively correlated edges were selected. Here stability of edge selection regardless

of restA versus restB FC data used is important. **(C)** Strength in the positive and negative restA masks were computed using restA FCs. Strengths were used as regressors in a linear model predicting the outcome measure. Here is important that the resulting model generalize to restB data from the same subjects. **(D)** Model generalizability to validation data was assessed. Here, it is important that the final model is generalizable to external data.

### 2.5 Association of resting-state networks to cognitive outcomes

Final masks for each outcome measure were defined by edges that appeared in at least 95% of the folds. We used binomial tests ($\alpha$ = 0.01) for each outcome measure to assess whether specific RSNs (e.g. DMN-DMN), or their interactions (e.g. DMN-FP), were overrepresented in these masks beyond what could be expected from an equal number of edges chosen at random. Edges from overrepresented networks (or interactions) were visualized using BrainNet viewer [55].

## 3 Results

### 3.1 Differential Identifiability

For both training and testing cohorts, $I_{diff}$ peaked at 41 PCs regardless of fold (Fig. 3, mean Training $I_{diff}$ = 67.09 & mean Validation $I_{diff}$ = 68.77, mean Training $I_{self}$ = 81.62 & Mean Validation $I_{self}$ = 81.68, mean Training $I_{others}$ = 32.46 & Mean Validation $I_{others}$ = 33.01, mean Training % variance explained = 71.86, mean Validation % variance explained = 71.80). We observed an almost two-fold increase in differential identifiability in the optimally reconstructed data (Fig 3 and Fig. S1. Such increase in whole-brain differential identifiability also increased the fingerprint at the functional edge level, as shown when using mean ICC (Fig. S2).

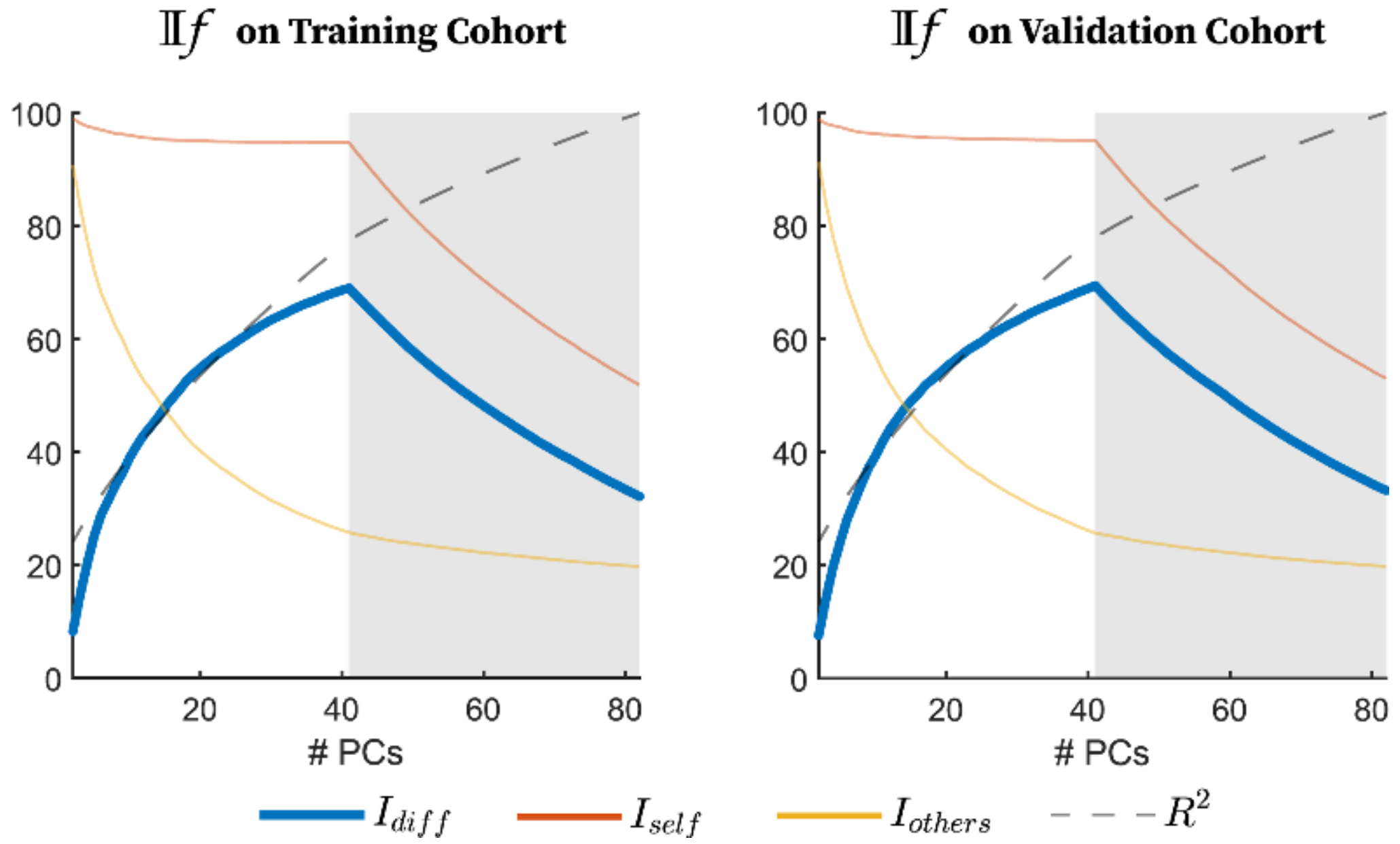

**Figure 3.** Mean of $\mathbb{I}f$ assessments on Training Cohort **(left)** and Testing Cohort **(right)**. Connectome level identifiability assessment. $I_{self}$ and $I_{others}$ represent similarity between test and retest FCs of the same vs different subjects, respectively, across number of PCs used for reconstruction. Differential identifiability ($I_{diff}$) is the difference between $I_{self}$ and $I_{others}$. The cumulative percent explained variance (100*$R^2$) across number of PCs used for reconstruction is also included.

### 3.1.1 Edge Selection – Stability and Specificity

Stability in edge selection between restA and restB (Fig. 4, colored lines) exhibited an optimal and stable range between 10-41 PCs both in terms of correlations and resulting selected edges associated to each outcome. The Frobenius norm of the edgewise correlation associated to each outcome measure (Fig.4 left) exhibited stable range of minimal divergence between RestA and RestB (10 to 41 PCs) after which divergence began to monotonically increase for all outcome measures. Overlap (Fig.4 right) between edges selected in RestA and RestB exhibited an optimal range of overlap (68% Clock Score – 76% Trails B) in the range of 10-41 PCs, then monotonically decreased after 41 PCs for all outcome measures. Specificity of edge selection, measured as pairwise similarity across outcomes (Fig.4, black lines), was stable across PCs for both the Frobenius norm and mask overlap. It is critical to highlight that all PCs were used for reconstruction (i.e. using original FC data) restA-restB similarity (colored lines) for any given outcome approachd similarity across outcomes of restA masks (black line).

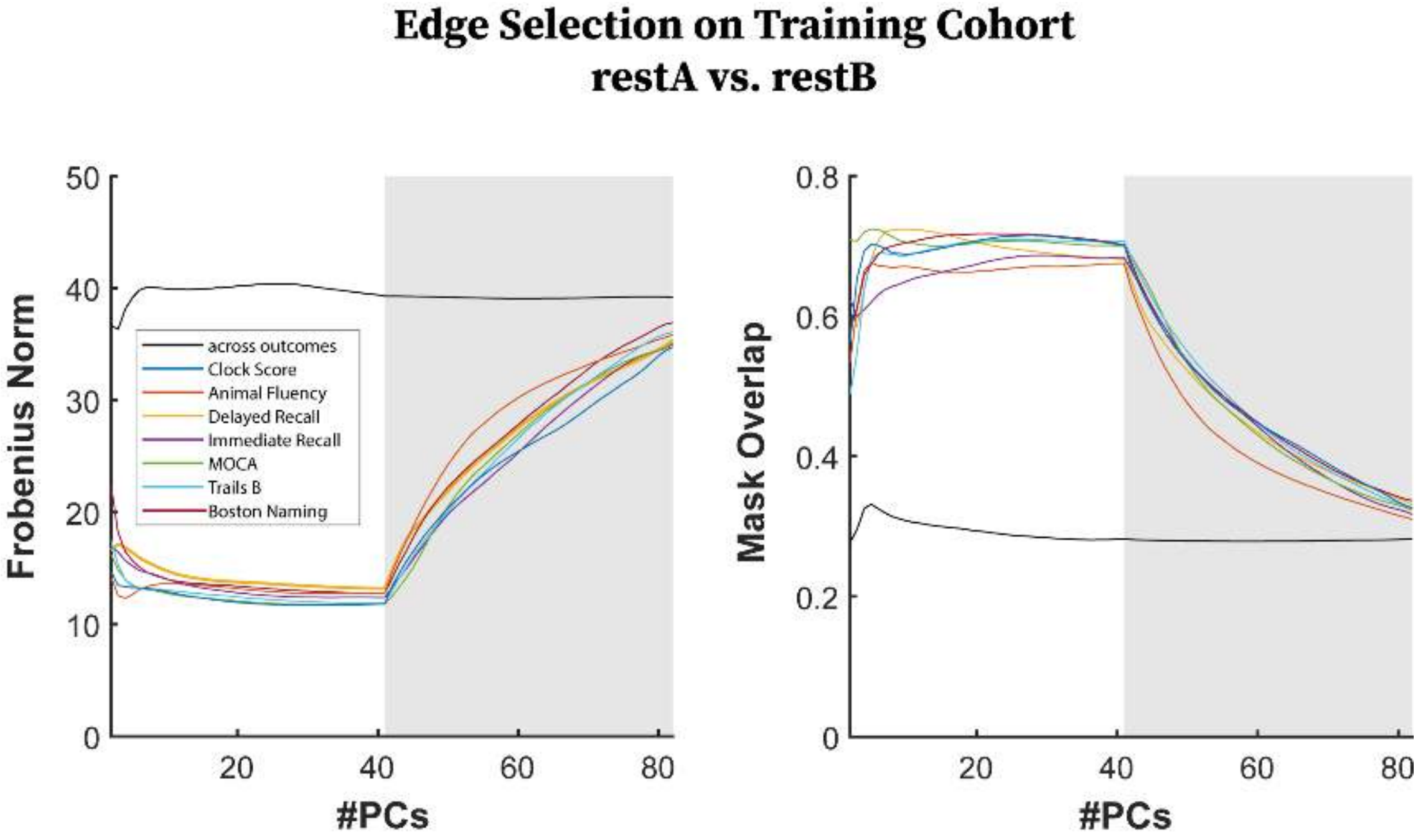

**Figure 4. (Left)** (Colored Lines) Frobenius norm of correlation matrices associated to each outcome measure for restA FCs versus restB FCs. (Black line) Average pairwise Frobenius Norm of correlation matrices between two different outcome measures using only restA FCs. **(Right)** (Colored Lines) Mask overlap between restA FCs versus RestB FCs, for each outcome measure. (Black Line) Average pairwise mask overlap between two different outcomes using only restA FCs.

### 3.1.2 Training Data: Test-Retest Generalizability

At the model fitting step, the performance of models built on restA data was evaluated on restA and restB FCs from the same subjects. For restA connectomes, on which the models were built, correlation between the predicted and estimated outcomes increased as the number of PCs increased, though at a slower rate after the optimal reconstruction point for $I_{diff}$ (Fig.5 left). In contrast, correlation between the predicted and estimated outcomes peaked at the optimal reconstruction point for $I_{diff}$ in the training data (41 PCs) when the model was applied on restB connectomes (Fig.5 right) and slightly decayed after 41 PCs.

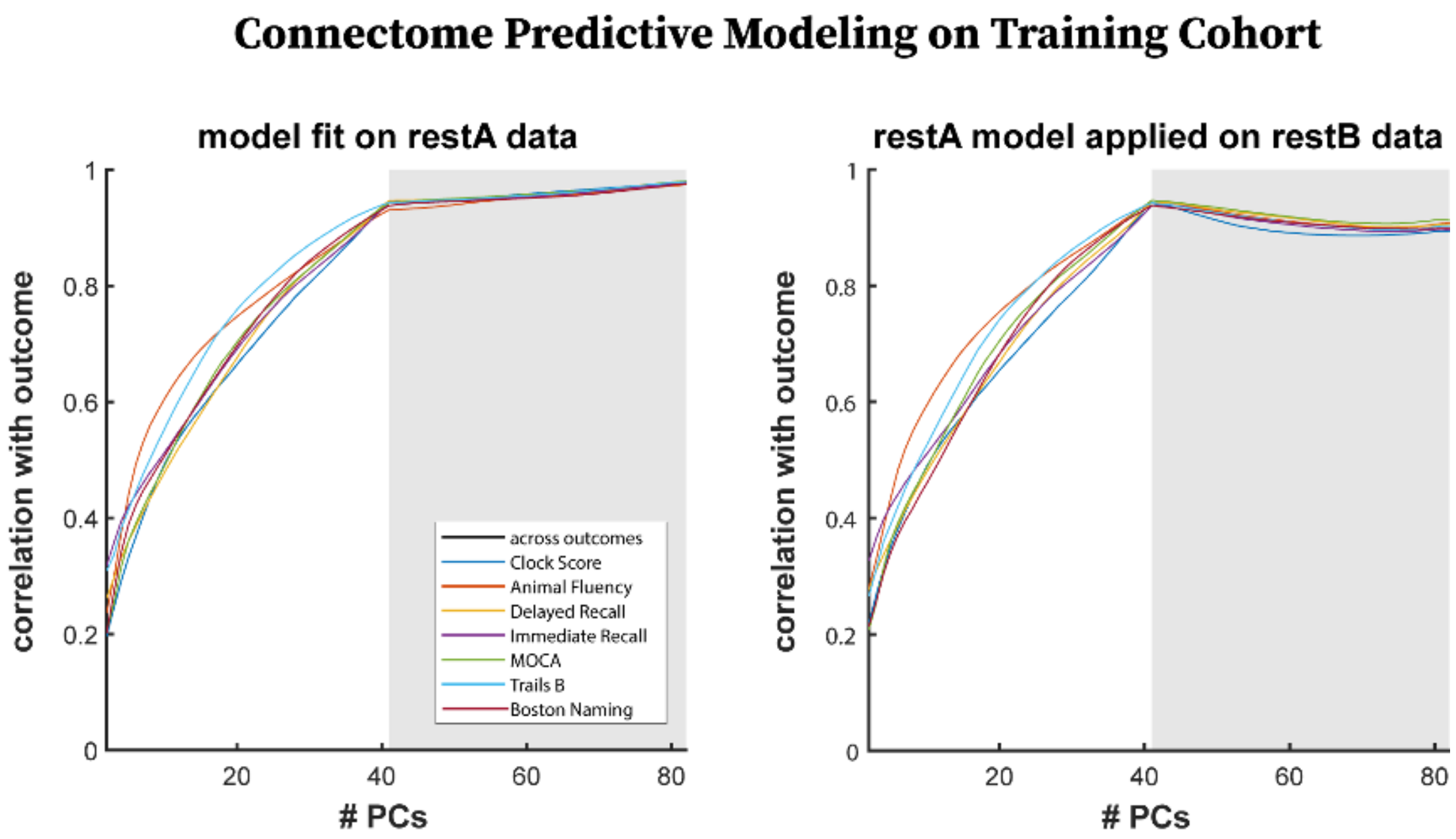

**Figure 5.** For all plots, restA Training FCs were used for edge selection and model fitting. **(Left)** Correlation between estimated and expected outcomes from models fit using restA FCs. **(Right)** Correlation between estimated and expected outcomes when models fit on restA Training FCs were applied to restB FCs from the same subjects.

### 3.1.3 Validation Cohort: Generalizability

Model performance was measured as mean correlation between the estimated and predicted outcomes across folds. Model performance from original FCs from the validation cohort ranged from 0.06 (±0.13) to 0.22 (±0.12) across outcomes (Fig6 left). Model performance from optimally reconstructed FCs from the validation cohort ranged from 0.20 (±0.17) to 0.33 (±0.10) across outcomes (Fig. 6 right). Model performance on validation data was significantly higher for optimally reconstructed FCs versus original FCs in 5 out of the 7 outcomes (Figure 6 right) and was not significantly different for the other two.

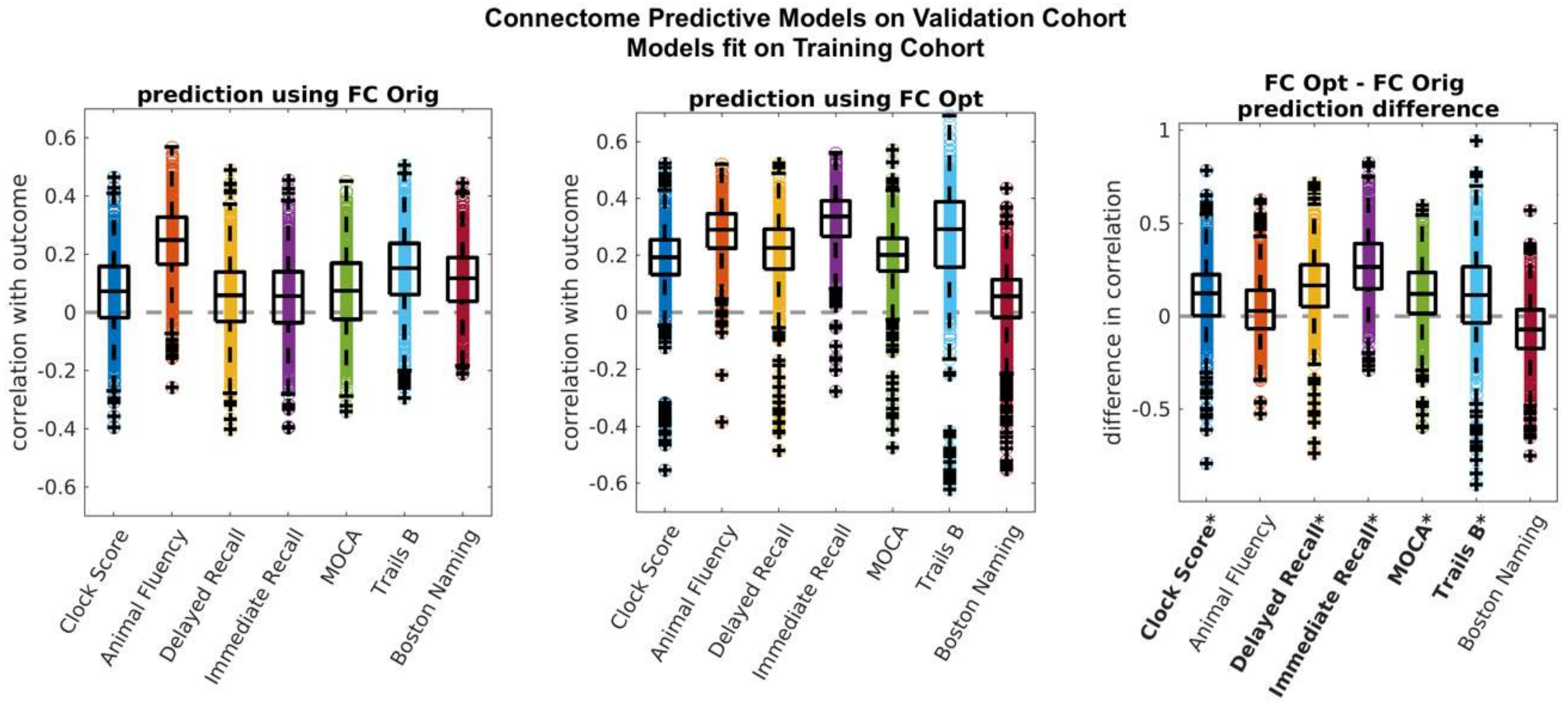

**Figure 6.** Model performance in validation cohort across 200 folds . Asterisk indicates outcomes for which performance was significantly improved in optimally reconstructed FCs versus original FCs (paired t-test, $\alpha$ = 0.01). The center line of each box corresponds to the median and the bounds to the $25^{th}$ and $75^{th}$ percentiles. Outliers are defined using 1.5*inter quartile range. **(Left)** Correlation between estimated and expected outcomes in original FCs from the validation

cohort. Models were fit using original FCs from the training cohort. **(Right)** Correlation between estimated and expected outcomes in optimally reconstructed FCs from the validation cohort. Models were fit on optimally reconstructed FCs from the training cohort.

### 3.1.4 Association of Resting State Networks to Cognitive Outcomes

We identified RSN components playing significant roles in prediction of each cognitive outcome and then assessed patterns in RSNs involved across cognitive outcomes. A total of 30 out of 45 possible RSN interactions were significantly over-represented, in either the positive or negative mask, across outcomes. The EC and SM networks played the largest role across outcomes participating in all 14 possible masks (7positive, 7 negative). The SAL network participated in 12 masks; the DA, DMN, and VIS networks participated in 11 masks; SUB and CER networks in 10 masks, and finally the L network in 7 masks. Within network connections had a significant role in 6 masks. Within DMN connectivity was negatively associated to performance in delayed recall and positively associated to trails B (note a smaller score in trails B denotes better performance). Within EC network connectivity was negatively associated to performance in MOCA and clock drawing. Within VIS connectivity was negatively associated to performance in the Boston naming test. Finally, within CER connectivity was associated to performance in trails B. Between network connectivity played a role in all 14 masks. Several RSNs played prominent roles in specific outcomes. Between network connections involving the VIS and SAL networks participated in 4 out of 9 RSN interactions associated to animal fluency. Connections involving the EC network played a role in 4 out of 6 RSN interactions associated to clock drawing. Finally, connections involving the CER played a role in 4 out of 9 RSN interactions associated to trails B. Several RSN interactions were present across outcomes. The DMN, EC, and SM networks participated in over-represented RSN interactions, either positively or negatively, across all outcomes (Fig. 7, Fig.S1-S6). SAL-EC connectivity was negatively associated with performance on MOCA (Fig. S3), AVLT immediate/delayed recall (Fig. S4-5), and trails B (Fig. S8). DA-SAL connectivity was positively associated with performance on MOCA (Fig. S3), animal fluency (Fig. 7), and Boston naming (Fig. S6). VIS-DA was associated negatively with animal fluency (Fig. 7) and trails B (Fig. S8). VIS-SM was associated positively with performance on both AVLT measures (Fig. S4-5) and animal fluency (Fig. 7). In contrast, SM-SUB connectivity was negatively associated with performance on these outcomes as well as clock drawing. VIS-DMN connectivity was negatively associated with both animal fluency and Boston naming. L-CER and DA-CER connectivity were associated positively with performance on MOCA (Fig. S3) and trails B (Fig. S8). Finally, DA-SAL connectivity was positively associated with performance on MOCA (Fig. S3) and Boston naming (Fig. S8).

**Table 4.** Significantly Overrepresented Resting State Networks for each outcome measure. RSNs (e.g. DMN-DMN) or their interactions (e.g. DMN-FP) represented above chance in edge selection (binomial test, $\alpha$ = 0.01). RSN abbreviations: visual (VIS), somato-motor (SM), dorsal attention (DA), Salience, ventral attention (SAL), limbic (L), executive control/fronto-parietal (EC), and default mode network (DMN), subcortical (SUB), and cerebellar (CER) network.

| | **Significant Resting State Networks** | |
|---|---|---|
| **Outcome Measure** | **Positive Mask** | **Negative Mask** |
| **MOCA** | DA-SAL<br>L-FP<br>L-CER | SM-L<br>SM-SUB<br>DA-DMN<br>DA-CER<br>SAL-EC<br>EC-EC |
| **Auditory Learning Immediate Recall** | VIS-SM<br>SUB-CER | SM-DA<br>SAL-EC<br>SAL-DMN |
| **Auditory Learning Delayed Recall** | VIS-SM<br>EC-CER | SAL-EC<br>DMN-DMN |
| **Boston Naming** | VIS-VIS<br>DA-SAL<br>L-EC<br>L-DMN<br>EC-SUB<br>DMN-SUB | VIS-SAL<br>VIS-DMN<br>SM-SUB<br>SM-CER |
| **Animal Fluency** | VIS-SM<br>VIS-SAL<br>DA-SAL<br>SAL-CER<br>L-EC | VIS-DA<br>VIS-DMN<br>SM-SAL<br>SM-SUB |
| **Clock Drawing** | SM-EC<br>EC-SUB | VIS-SUB<br>SM-SUB<br>EC-EC<br>EC-DMN |
| **Trail Making B** | VIS-DA<br>SM-SUB<br>DA-CER<br>SAL-EC<br>DMN-CER | SM-SAL<br>L-CER<br>DMN-DMN<br>CER-CER |

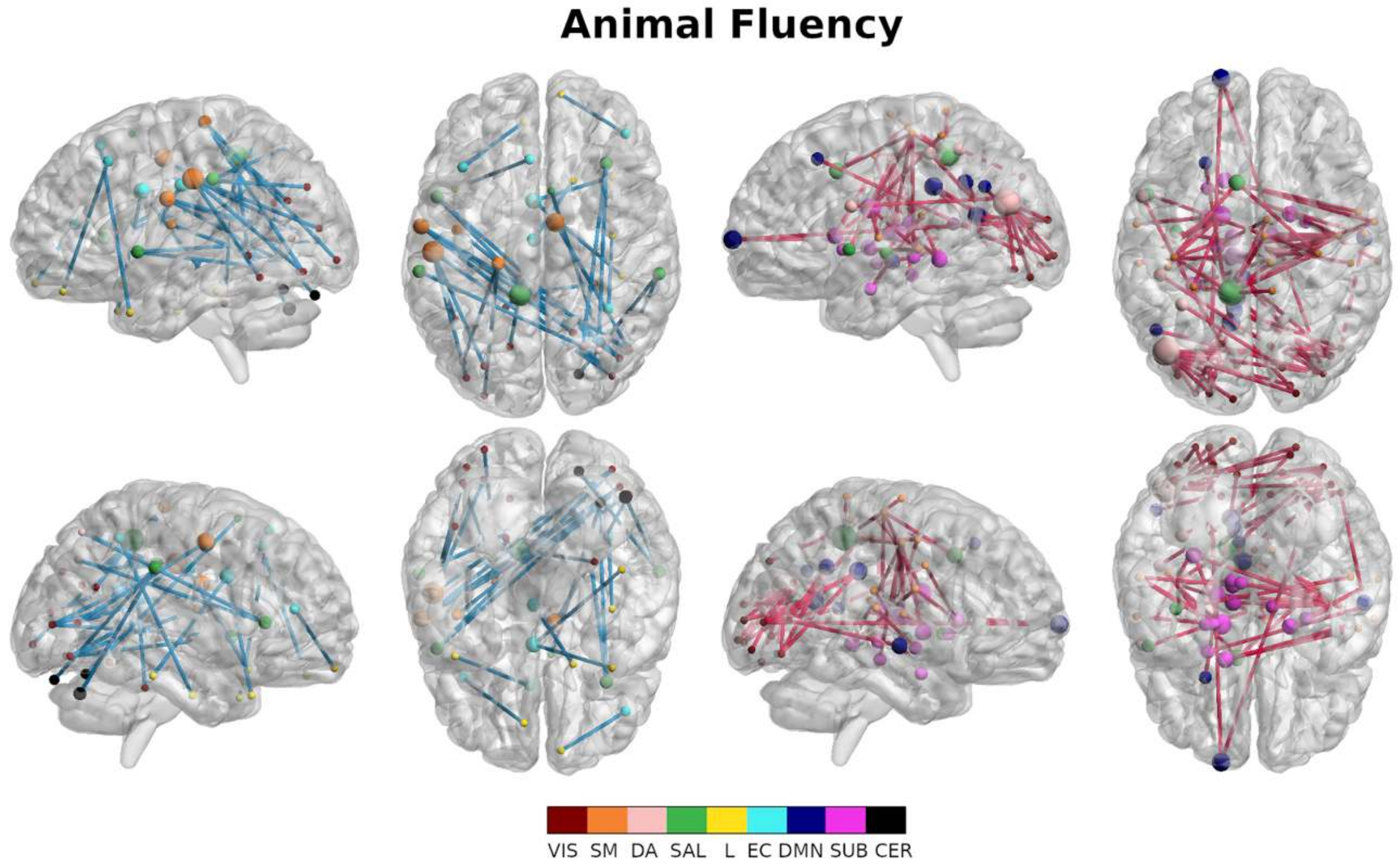

**Fig. 7:** Overrepresented edges (binomial test, $\alpha$ = 0.01) for the Animal Fluency Test. Positively associated edges (**left**) and negatively associated edges (**right**) are visualized separately. Nodes are sized according to their degree and colored according to resting state network membership. Positive mask edges are colored blue while negative mask edges are colored red.

## 4 Discussion

Our work provides a comprehensive whole brain and whole cognitive spectrum view on the relationship between resting-state functional connectivity and cognition in AD and makes progress towards making subject level predictions of cognition from FC biomarkers. We accomplished that by improving the robustness of connectome predictive models of AD using $\mathbb{I}f$, which improved test/retest generalizability of these models and allowed for significantly improved predictions of cognition from external FC data for five out of the seven outcomes evaluated. Finally, robust edge selection allowed for identification of RSN motifs associated with cognitive deficits in AD.

### 4.1 Differential Identifiability

The use of FC as a biomarker in clinical settings requires major advancements in subject level identifiability of FC. In this work, we improve subject level FC identifiability, as measured using differential identifiability, using group level PCA. As demonstrated by other datasets [12], the number of PCs necessary to optimize differential identifiability corresponded to the number of subjects in the cohort (Fig. 3 Blue Line). This indicates that, while the dimensionality of the input data is twice the number of subjects (due to inclusion of test and retest data), the subject dimensionality of the data is the cutoff for a more accurate representation of individual functional connectomes. Also, as shown previously [12], optimizing $I_{diff}$, a coarse whole brain measure (Fig. 3 blue line showing Idiff & Fig. S1?), also robustly increased test-retest reliability at the level of individual edges (Fig. S2). Optimal reconstruction retained 80% of the variance in the data (Fig. 3 purple line), indicating that around 20% of variance present in individual FC estimates is not representative of robust individual characteristics, despite the extensive preprocessing of BOLD time series used here and described in detail in [32]. Additionally, we note that $I_{diff}$ for this dataset is much higher than what we saw in previous data where $I_{diff}$ was optimized by splitting the resting state time series in half [12], highlighting that datasets with coarse temporal acquisition or datasets including clinical populations may benefit greatly from this group level PCA cleaning technique in order to improve individual level estimates of FC.

### 4.2 Effect of Differential Identifiability on Connectome Predictive Modeling

When assessing clinical populations with CPM, one of the ultimate goals is to identify critical functional subcircuits associated with specific cognitive deficits. Therefore, a minimum criterion that must be met is that edge selection should be robust between test/retest data (e.g. fMRI runs or sessions) from the same subjects. Thus, we took advantage of previous splitting of fMRI data into restA and restB for purposes of uncovering connectome fingerprinting to compare edge selection performed separately on for restA vs. restB FCs. Using $\mathbb{I}f$, we were able to improve the robustness of CPM in identifying functional subcircuits associated to specific cognitive deficits. Stability of edge selection displayed an optimal regime (12-41 PCs), after which it exponentially worsened for all outcome measures (Fig. 4; see colored lines). Overlap between restA and restB edge selection (Fig. 4) for optimally reconstructed data increased by an average of 30% from raw data, with an average peak overlap of 65% across outcome measures.

$\mathbb{I}f$ did not affect the relative specificity in edge selection across outcome measures (Fig. 4 black lines). Frobenius norm between outcomes remained constant around 40 and mask overlap remained constant at around 30%. This implies that the "distance" between mappings of different outcomes is preserved across PCs whereas the distance between restA-restB mappings for a single outcome is reduced as we move from original FCs to optimally reconstructed FCs (1/2 total number of PCs). It is noteworthy that for original FCs (equivalent to reconstructing with all PCs) restA-restB overlap approached overlap across outcomes. This implies that mappings of a single outcome based on two sessions of FCs of the same subjects are as non-specific as the mappings of different outcomes using a single session of FC. From a clinical standpoint, where understanding which brain systems are affected is as important as predicting cognitive outcomes, this situation hampers the utility of the model. This situation is highly alleviated when performing the $\mathbb{I}f$ prior to CPM, where restA-restB Frobenius norm is significantly lower than across outcomes (Fig. 4 left) and restA-restB mask overlap is significantly higher than across outcomes (Fig. 4 right).

In addition to improving robustness of edge selection, we also modestly improved prediction of cognitive and behavioral outcomes in new FC data from the same subjects using $\mathbb{I}f$More importantly, the addition of a test/restest validation step to CPM showed that reconstructing FC at the optimal point for $I_{diff}$ reduces overfitting to the training data as evidenced by a continued increase in model performance after 41 PCs for restA data from Training subjects versus a decrease after 41 PC for restB data. However, as optimally reconstructed restA and restB FCs come from the same orthogonal bases, it could be argued that their independence is further reduced upon implementation of the $\mathbb{I}f$, thus the improved prediction. To ensure this was not the case, we showed that optimal reconstruction of FC improved the generalization of models from the training cohorts to the validation cohorts for most of the cognitive outcomes (Fig. 6). Note that the differential identifiability pipeline was run separately on the training and validation cohorts at each fold, thus fully maintaining the independence of training data and validation data.

### 4.3 Association of Resting State Networks to Cognitive Outcomes

Previous literature making predictions from ADNI fMRI data focused solely on prediction of global cognitive status [45] or diagnostic status [56]. In contrast, we assessed the involvement of RSN systems across the cognitive spectrum in AD and found several motifs consistent with previously reported literature about the role of RSNs in AD and in general cognition. We identified the EC network, DMN, and SM networks as globally associated to all outcomes. The central role of the EC and DMN in AD [42, 57-59] and their strong associations with amyloid [42, 60-62] and tau deposition [61, 63, 64] has been consistently documented. The sparing of the SM network from AD pathology has also been consistently documented [65]. Previous findings have also indicated compensatory connectivity occurring through the SM network [66], which may explain our findings of the prominent role of this network in predicting cognitive outcomes in AD. In addition to globally associated networks, we found networks that appeared consistently across outcomes associated with specific cognitive domains. For instance, the functional connectivity between the SAL and DA network as well as between the EC network and the SAL network were consistently associated with cognitive tasks that included a large attention component, consistent with a study showing that interaction between these networks plays a critical role in perceptual attention [67]. Impaired connectivity between these networks has also been previously associated with cognitive decline in AD measured by increasing clinical dementia rating score [57]. We also identified that interactions between the VIS network and other RSNs were consistently associated with tasks that required item generation in the context of verbal memory retrieval (i.e., AVLT, MOCA) or spontaneous generation of items belonging to a given category (i.e., animal fluency). This finding suggests an interactive role of the visual system with other functional subcircuits when executing tasks requiring semantic organization and imagery. This role of the visual system is supported by other studies identifying activation of the visual cortex and cognitive networks in imagery and semantic association tasks [68, 69]. Additionally, the visual cortex has also been implicated in visual short term memory and working memory [68]. Furthermore, in AD, connectivity of the visual system has been previously associated to neurofibrillary tangle deposition [64] and with cognitive complaints in cognitively normal or MCI subjects [16]. We consistently saw a significant role of cerebellar connectivity in tasks with significant motor components. Intra-cerebellar network connectivity in trails, while between network cerebellar connections with the limbic system and with the dorsal attention network were also associated with performance on trails B as well as MOCA. The trails B task is a subset of the MOCA battery [70]. Thus findings of RSN associations, consistent with previously reported roles of these RSNs in AD and cognition indicates that our unified framework not only produces robust prediction, but also anatomically relevant mapping of cognitive deficits to resting state functional connectivity.

### 4.4 Limitations and Future work

The unified identifiability-CPM framework proposed here provides many opportunities for improving the clinical utility of FC. However, an important and necessary step to improve the clinical utility of FC is to evaluate results obtained using this unified framework on a completely external dataset such as ADNI3, which includes similar acquisition from different scanner types. This will require the estimation of final hyper parameters from the ensemble of those estimated here using the entire ADNI2 cohort and the edges appearing in the final masks. In addition to external validation of the framework, our results indicate that there are other opportunities to improve both edge selection and predictive capability of FC. Despite showing significant improvement in robustness of edge selection using our framework, we were still under 80% test/retest overlap in edge selection for all outcome measures. Edge selection may potentially be improved by taking into account the network relationship between edges, as opposed to using edgewise correlation with thresholding which treats edges as univariate independent entities. This has been previously done using methods such as Partial Least Squares regression [29]. One could also incorporate concepts from the Network Based Statistics framework to control for spurious, small connected components [71]. Controlling for such components would allow the edge selection step to be thought of as the identification of the functional subcircuits associated with a given outcome, enabling the use of network science measurements [4-6] which may provide additional predictive power and provide further insight into the mechanisms of cognition and behavior. Thus, incorporating such methodologies may provide additional improvements in prediction to those shown here. As within and between network connections tend to have different properties [10], another avenue to test could be estimating separate masks and coefficients for within and between RSN connections. Finally, CPM may also prove useful in predicting change in cognitive outcomes over time [72], thus assessing the effect of differential identifiability on connectome predictive of longitudinal outcomes in AD would be a worthy contribution towards improving FC utility as a clinical biomarker.

## 5 Conclusions

Our framework improved the robustness of individual level prediction of cognition from FC, which is the first step towards clinical use of FC and better understanding of how functional connectivity supports cognition in AD. We showed that the joint framework of differential identifiability with connectome predictive modeling improves the quality of models obtained from CPM in terms of stability of edge selection, test/retest generalizability, and generalizability to external data. Additionally, we showed that the use of two FC sessions from each subject provides a unique perspective when assessing and validating connectome predictive models. Finally, improving the robustness of edge selection allowed for reliable assessment of the associations between functional connectivity and cognitive deficits in AD. Our findings indicate both specific and global associations of resting state functional connectivity with cognitive deficits in AD which are consistent with previous literature regarding the roles resting state networks play in both cognition and AD.

**Funding:** This work was supported by the National Institutes of Health NIA F32AG062157, NIA R56, AG057195/U01 AG057195, NIA R01AG057739, NIA R01 AG040770, NIA K02 AG048240, NIA P30 AG010133, NIA U01 AG02490, NIH R01EB022574, NIH R01MH108467, the Baekgaard family, the Indiana Alcohol Research Center P60AA07611, and the Purdue Discovery Park Data Science Award "Fingerprints of the Human Brain: A Data Science Perspective".

**Acknowledgements:** Data collection and sharing for this project was funded by the Alzheimer's Disease Neuroimaging Initiative (ADNI) (National Institutes of Health Grant U01 AG024904) and DOD ADNI (Department of Defense award number W81XWH-12-2-0012). ADNI is funded by the National Institute on Aging, the National Institute of Biomedical Imaging and Bioengineering, and through generous contributions from the following: AbbVie, Alzheimer's Association; Alzheimer's Drug Discovery Foundation; Araclon Biotech; BioClinica, Inc.; Biogen; Bristol-Myers Squibb Company; CereSpir, Inc.; Cogstate; Eisai Inc.; Elan Pharmaceuticals, Inc.; Eli Lilly and Company; EuroImmun; F. Hoffmann-La Roche Ltd and its affiliated company Genentech, Inc.; Fujirebio; GE Healthcare; IXICO Ltd.; Janssen Alzheimer Immunotherapy Research & Development, LLC.; Johnson & Johnson Pharmaceutical Research & Development LLC.; Lumosity; Lundbeck; Merck & Co., Inc.; Meso Scale Diagnostics, LLC.; NeuroRx Research; Neurotrack Technologies; Novartis Pharmaceuticals Corporation; Pfizer Inc.; Piramal Imaging; Servier; Takeda Pharmaceutical Company; and Transition Therapeutics. The Canadian Institutes of Health Research is providing funds to support ADNI clinical sites in Canada. Private sector contributions are facilitated by the Foundation for the National Institutes of Health (www.fnih.org). The grantee organization is the Northern California Institute for Research and Education, and the study is coordinated by the Alzheimer's Therapeutic Research Institute at the University of Southern California. ADNI data are disseminated by the Laboratory for Neuro Imaging at the University of Southern California.

## Supplementary Figures

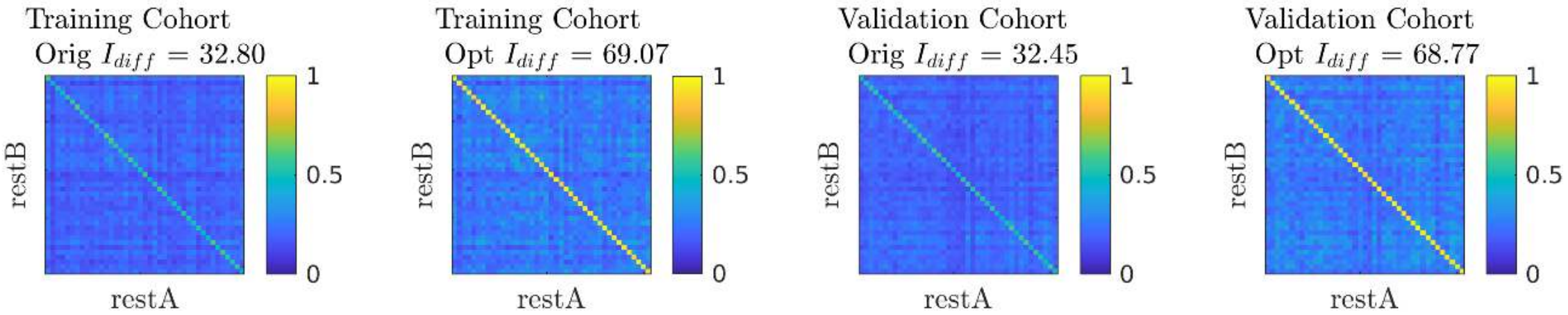

**Fig. S1:** Average Identifiability matrices over the 1000 folds for original FCs and FCs reconstructed at the optimal point for differential identifiability. (Left) Average matrices in the training cohorts. (Right) Average Matrices for the validation cohorts.

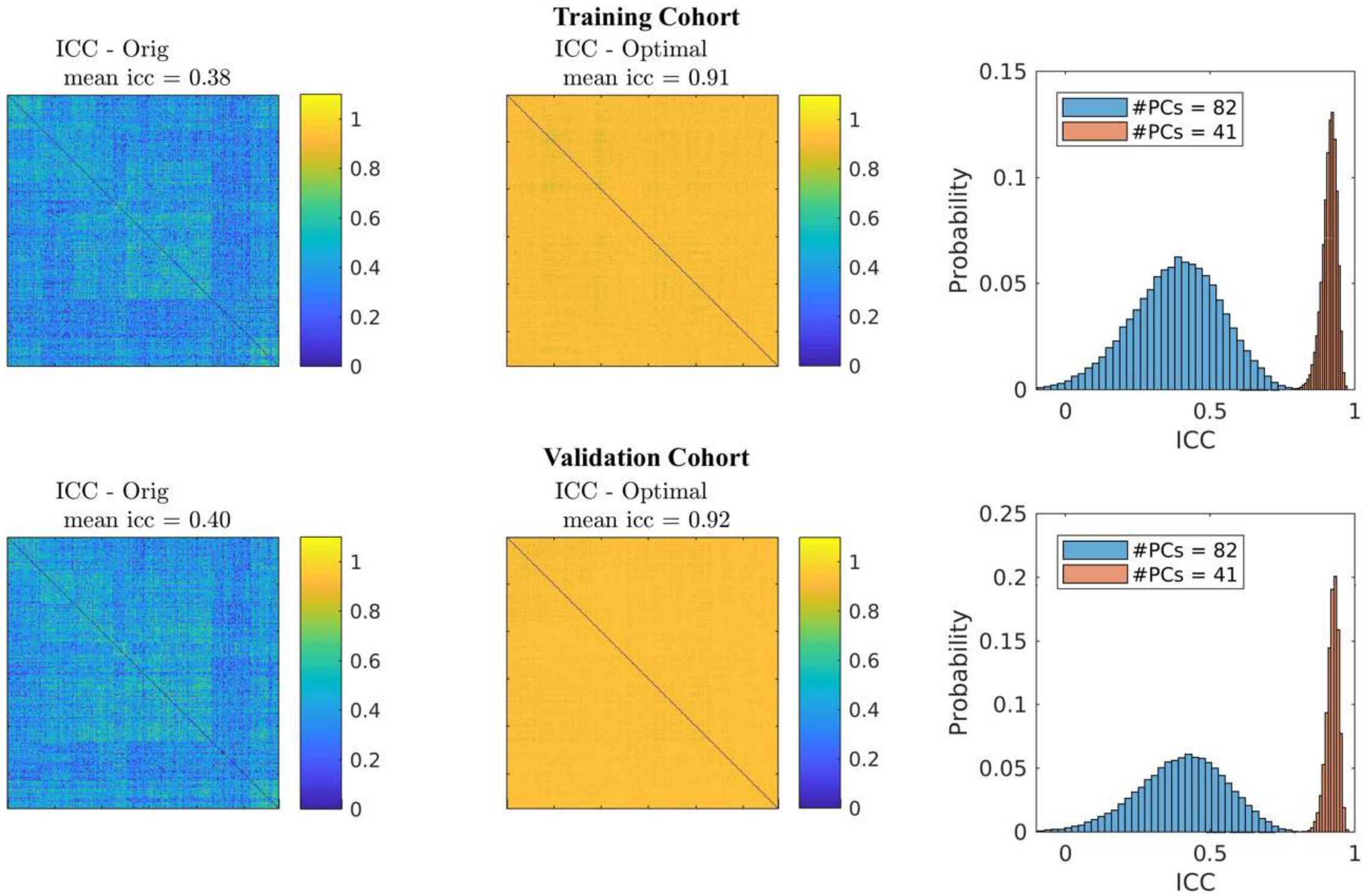

**Fig. S2:** Average edgewise ICC over 1000 folds for original FCs and FCs reconstructed at the optimal point for differential identifiability. Edges in ICC matrices are ordered according to RSN membership. (**left**) Edgewise ICC matrix for original FCs. (**Middle**) Edgewise ICC matrix for optimally reconstructed FCs. (**right**) Distribution of edgewise ICC values before for original FCs (blue, 82 PCs) and optimally reconstructed FCs (red, 41 PCs).

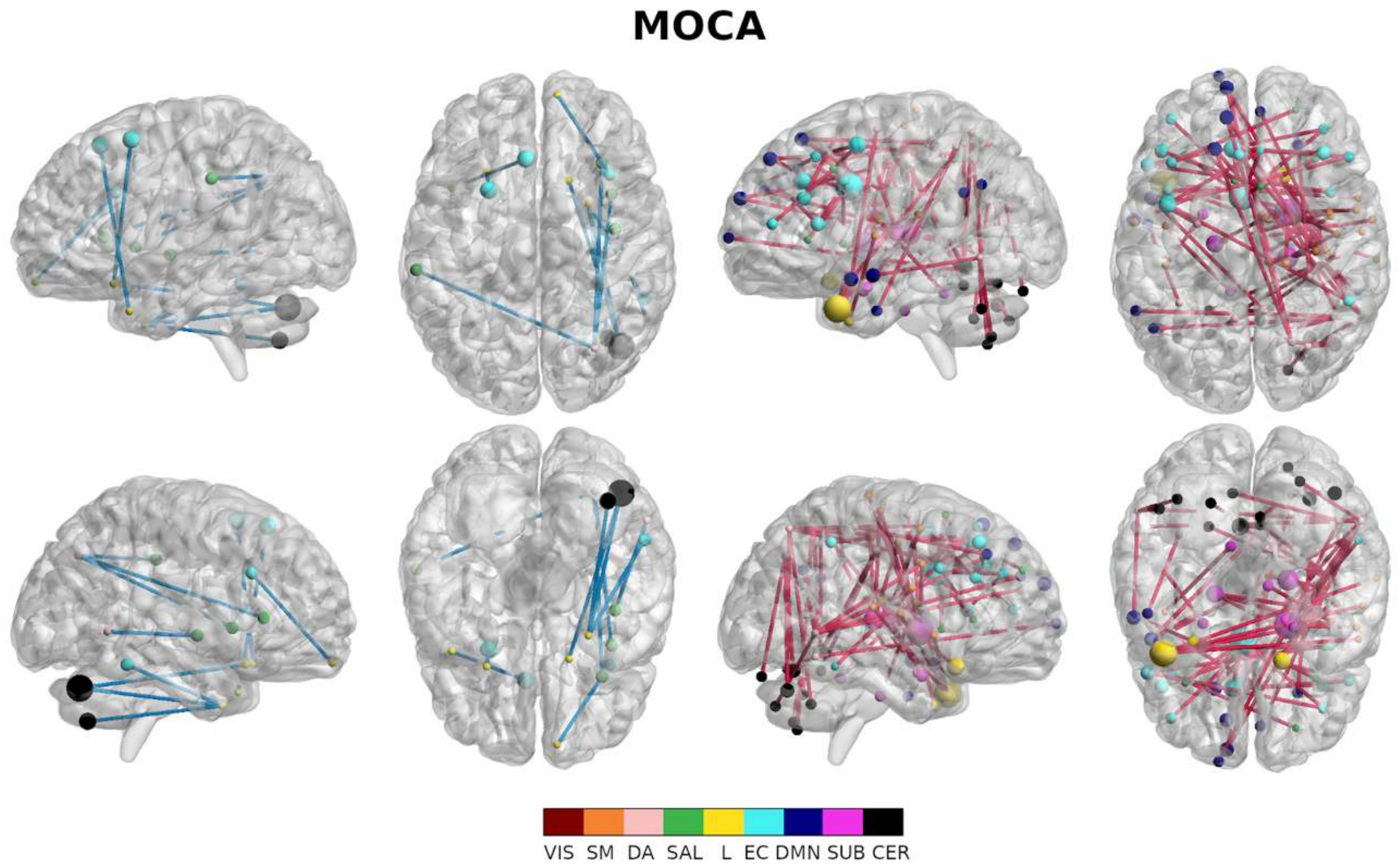

**Figure S3:** Over represented edges (binomial test, $\alpha$ = 0.01) for the Montreal Cognitive Association Test (MOCA). Positively associated edges (**left**) and negatively associated edges (**right**) are visualized separately. Nodes are sized according to their degree and colored according to resting state network membership. Positive mask edges are colored blue while negative mask edges are colored red.

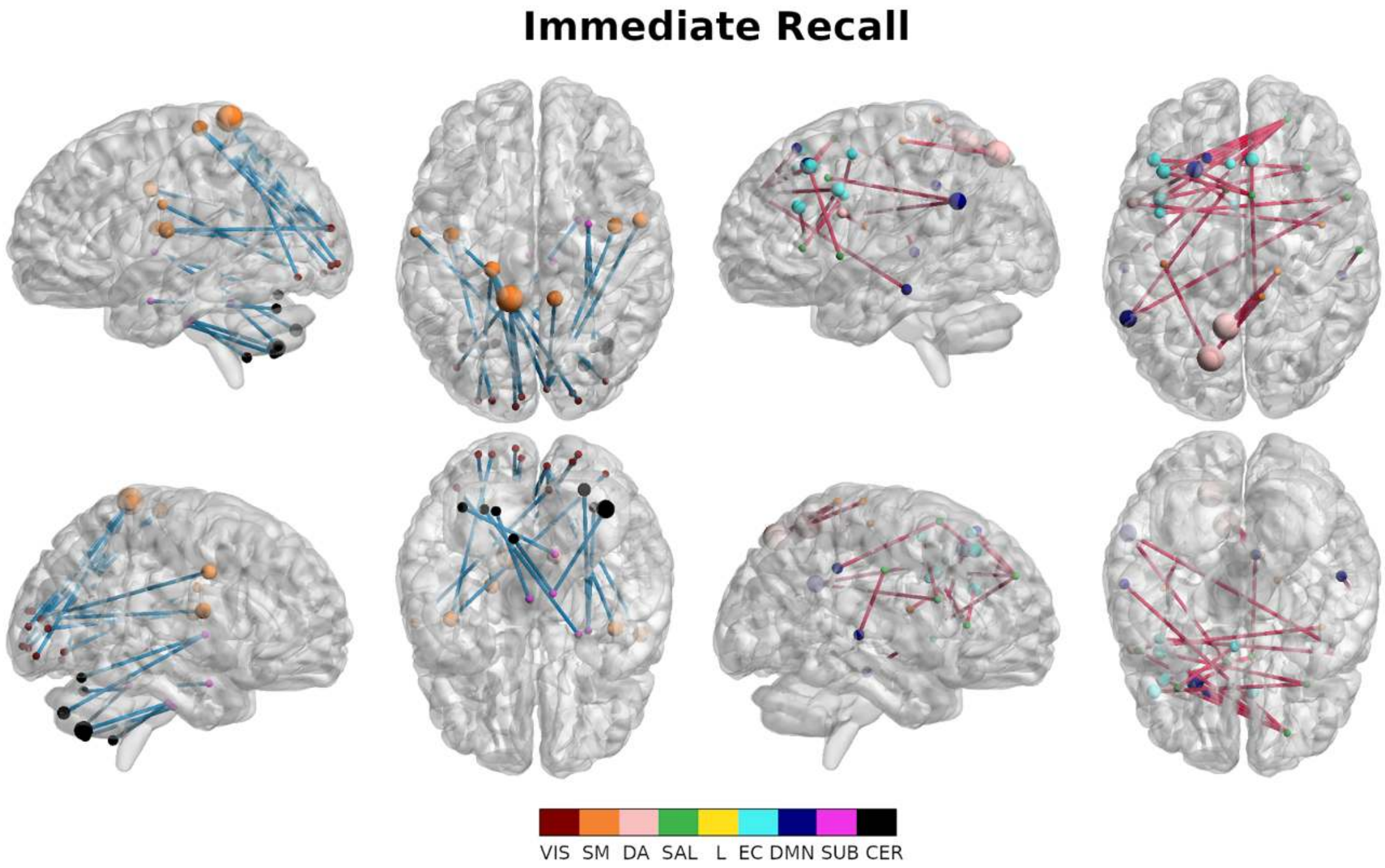

**Figure S4:** Over represented edges (binomial test, $\alpha$ = 0.01) for the AVLT Immediate Recall test. Positively associated edges (**left**) and negatively associated edges (**right**) are visualized separately. Nodes are sized according to their degree and colored according to resting state network membership. Positive mask edges are colored blue while negative mask edges are colored red.

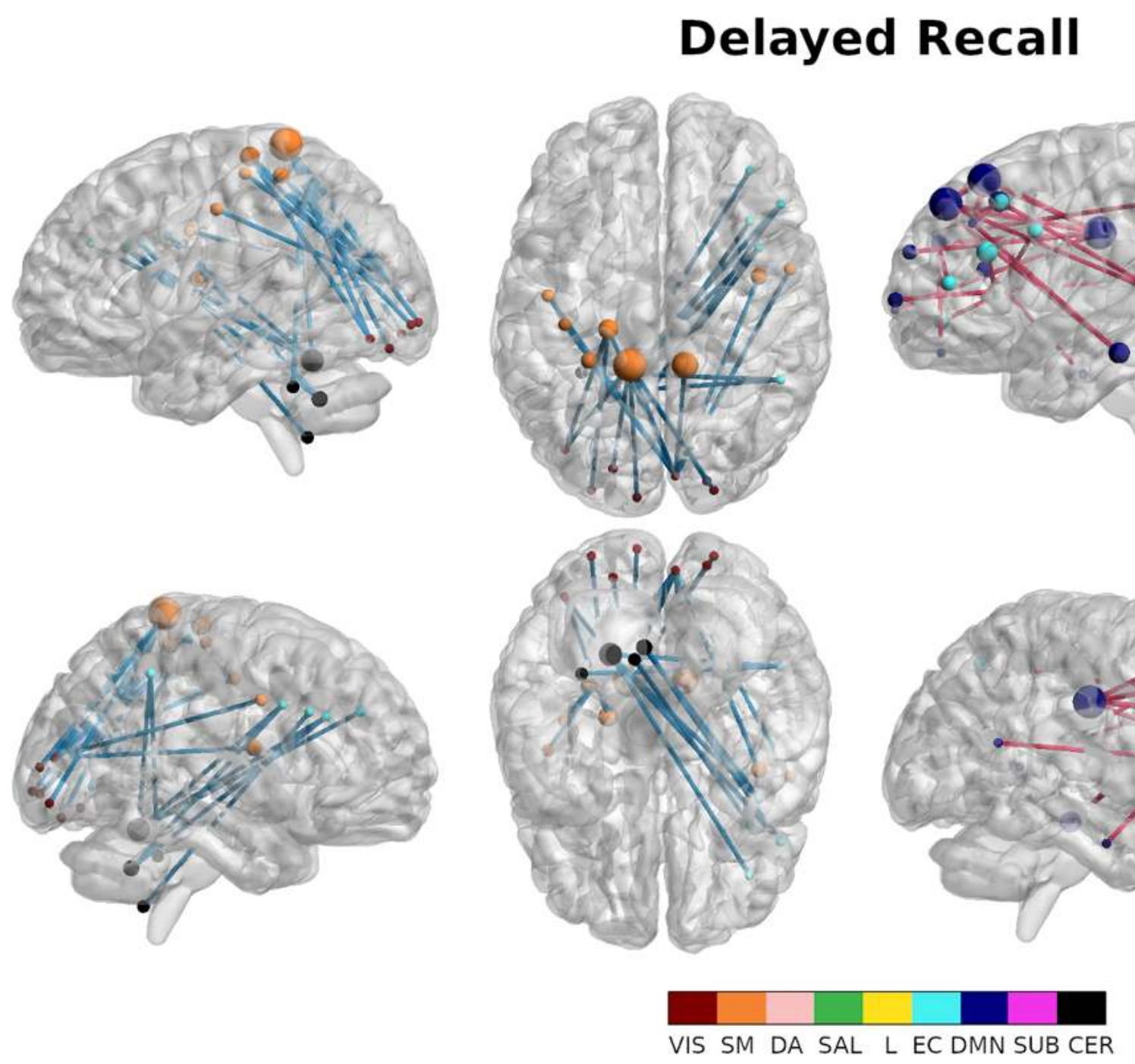

**Figure S5:** Over represented edges (binomial test, $\alpha$ = 0.01) for the AVLT Delayed Recall test. Positively associated edges (**left**) and negatively associated edges **(right**) are visualized separately. Nodes are sized according to their degree and colored according to resting state network membership. Positive mask edges are colored blue while negative mask edges are colored red.

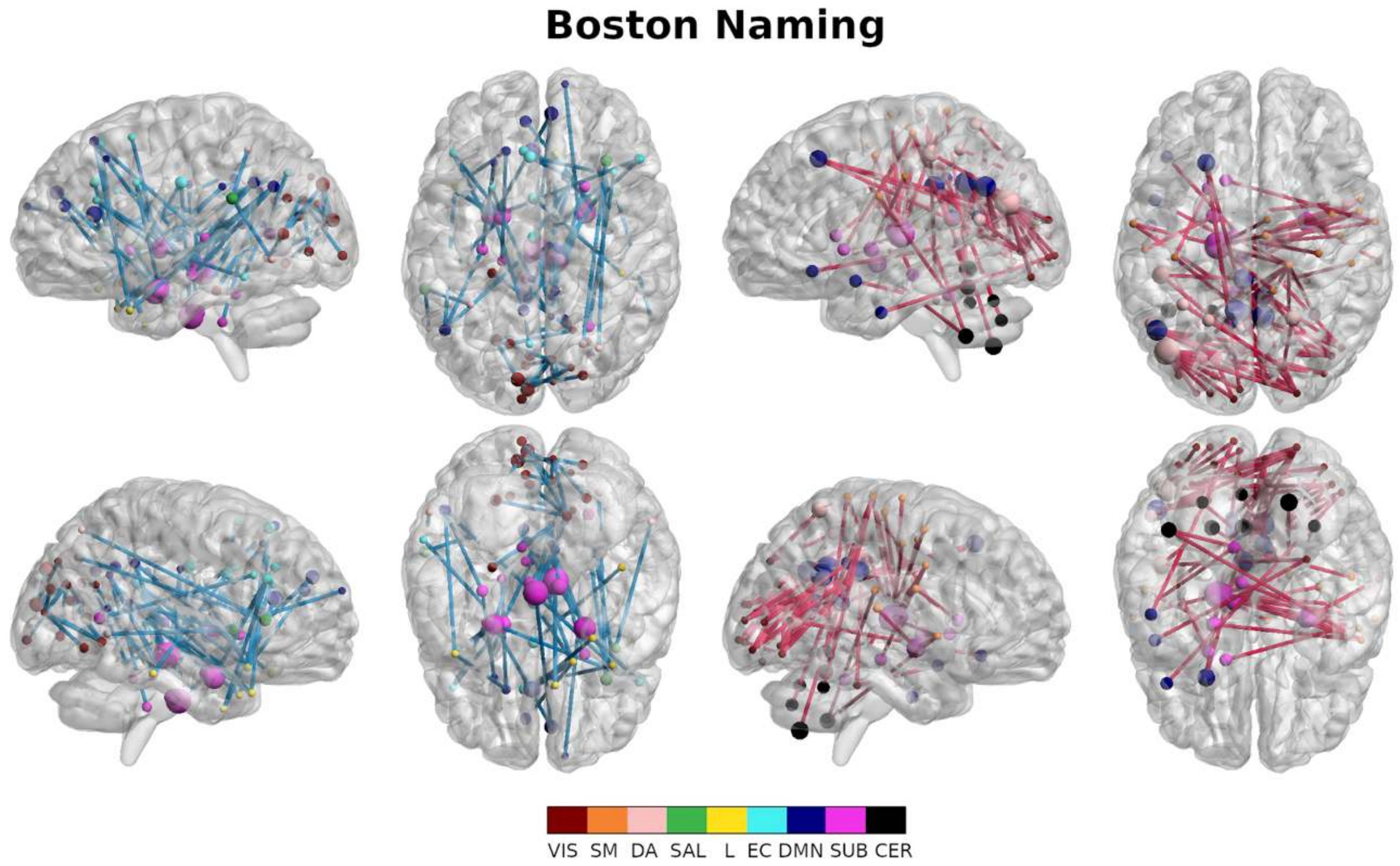

**Figure S6.** Over represented edges (binomial test, $\alpha$ = 0.01) for the Boston Naming Test. Positively associated edges (**left**) and negatively associated edges (**right**) are visualized separately. Nodes are sized according to their degree and colored according to resting state network membership. Positive mask edges are colored blue while negative mask edges are colored red.

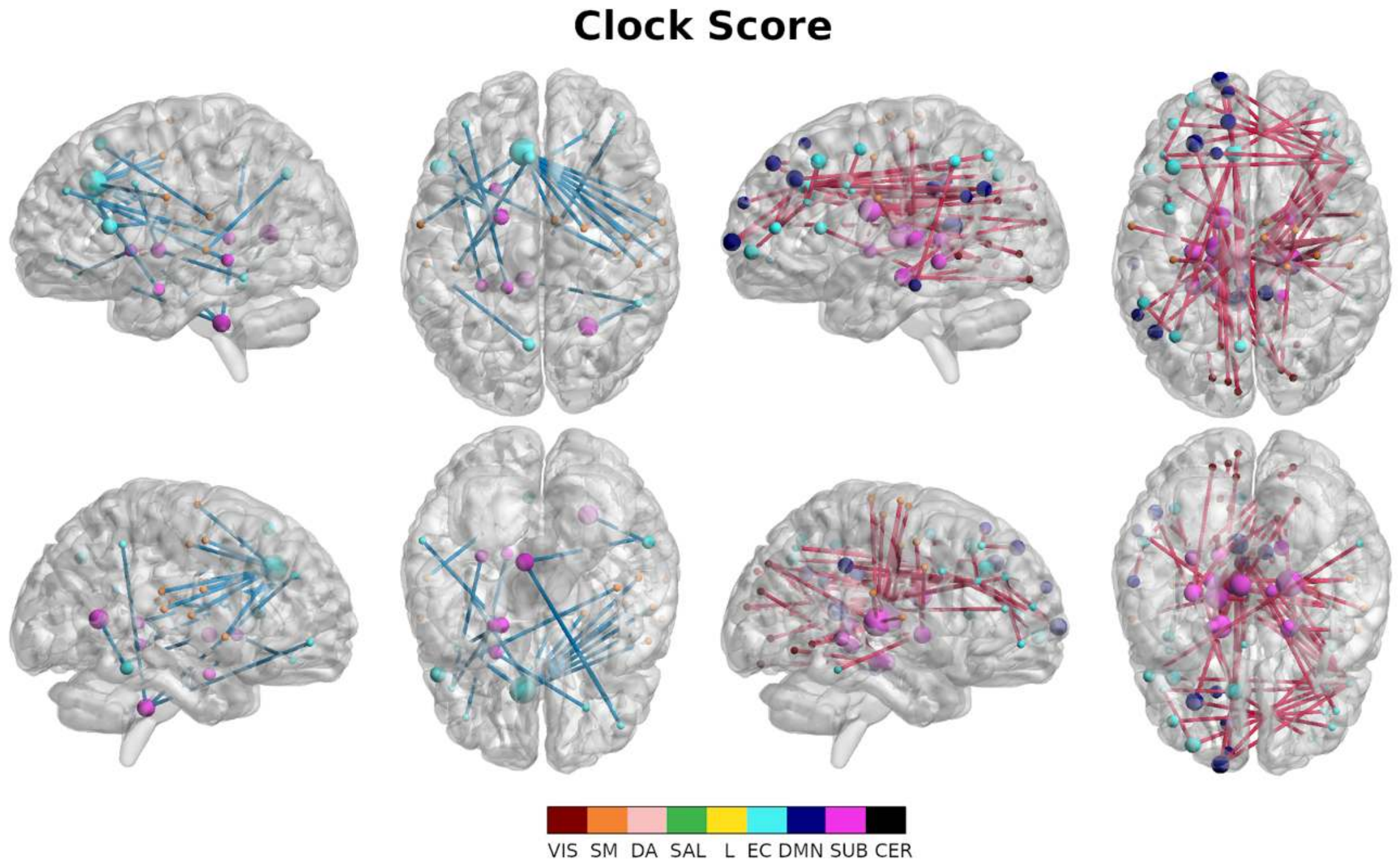

**Figure S7.** Over represented edges (binomial test, $\alpha = 0.01$) for the Clock Drawing Test. Positively associated edges (**left**) and negatively associated edges (**right**) are visualized separately. Nodes are sized according to their degree and colored according to resting state network membership. Positive mask edges are colored blue while negative mask edges are colored red.

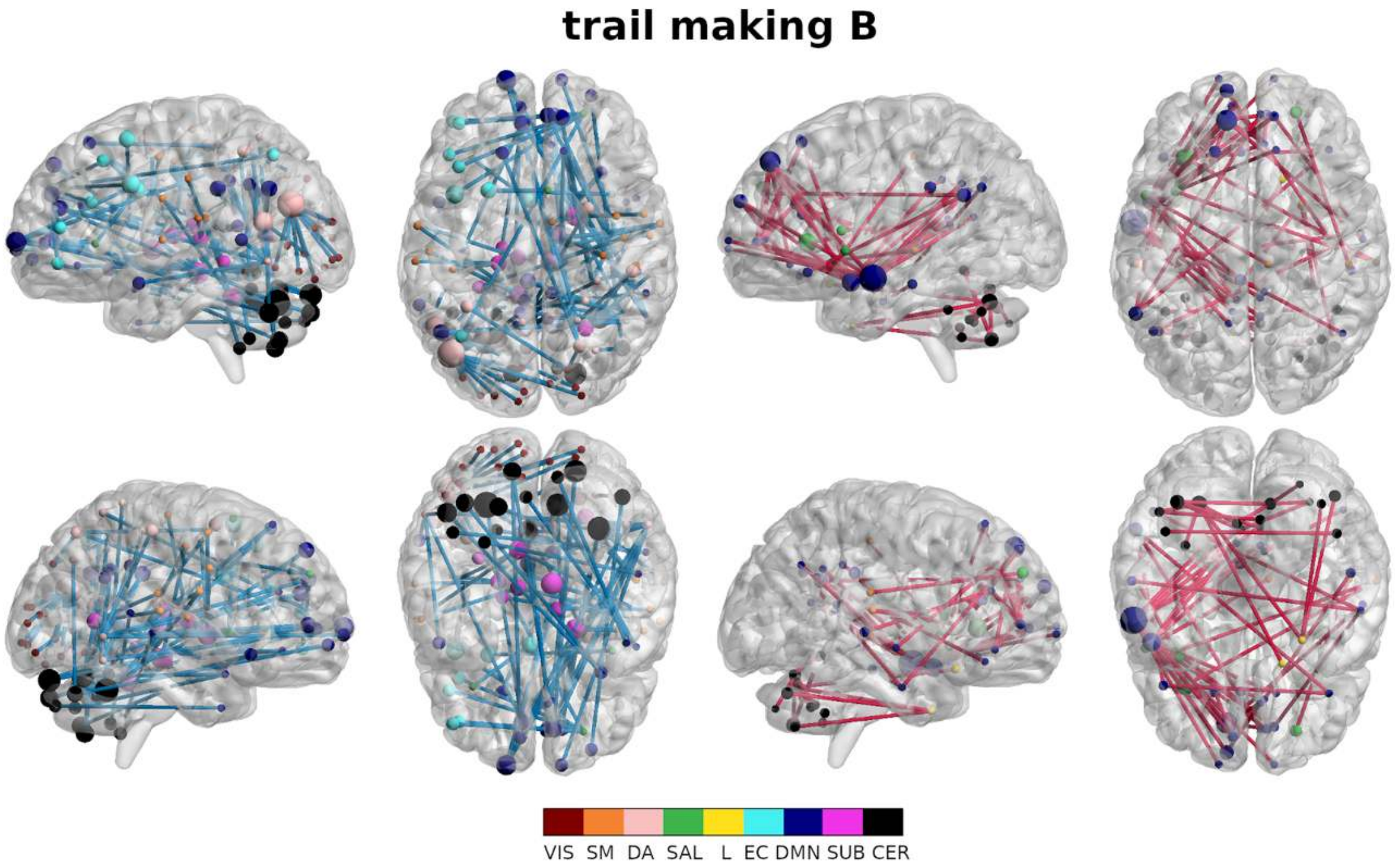

**Figure S8:** Over represented edges (binomial test, $\alpha$ = 0.01) for the Trail Making B test. Positively associated edges (**left**) and negatively associated edges (**right**) are visualized separately. Nodes are sized according to their degree and colored according to resting state network membership. Positive mask edges are colored blue while negative mask edges are colored red.